\documentclass{article}

\usepackage{arxiv}

\usepackage[utf8]{inputenc} 
\usepackage[T1]{fontenc}    
\usepackage{url}            
\usepackage{booktabs}       
\usepackage{amsfonts}       
\usepackage{nicefrac}       
\usepackage{microtype}      
\usepackage{lipsum}
\usepackage{float}
\usepackage{graphicx}
\graphicspath{ {./images/} }
\usepackage{xcolor}
\usepackage[capitalize]{cleveref}
\crefname{section}{Sec.}{Secs.}
\Crefname{section}{Section}{Sections}
\Crefname{table}{Table}{Tables}
\crefname{table}{Tab.}{Tabs.}

\title{Multimodal contrastive learning for diagnosing cardiovascular diseases from electrocardiography (ECG) signals and patient metadata}

\author{
\textbf{Tue M. Cao} \\ VinUni-Illinois Smart Health Center, VinUniversity  \\ Hanoi University of Science and Technology \\ \texttt{tuetuetue123123@gmail.com} \\
  \And
 \textbf{Nhat H. Tran} \\ VinUni-Illinois Smart Health Center, VinUniversity  \\ Hanoi University of Science and Technology \\ \texttt{tranhongnhat1911@gmail.com} \\
 \and
 \textbf{Phi Le Nguyen} \\ School of Information and Communication Technology, \\ Hanoi University of Science and Technology  \\ \texttt{lenp@soict.hust.edu.vn} \\
 \and
 \textbf{Hieu Pham} \\ VinUni-Illinois Smart Health Center, \\ College of Engineering \& Computer Science, VinUniversity  \\ Coordinated Science Laboratory, University of Illinois Urbana-Champaign \\ \texttt{hieu.ph@vinuni.edu.vn} \\
}

\begin{document}
\maketitle

\section{Introduction}

This work discusses the use of contrastive learning and deep learning for diagnosing cardiovascular diseases from electrocardiography (ECG) signals. While the ECG signals usually contain 12 leads (channels), many healthcare facilities and devices lack access to all these 12 leads. This raises the problem of how to use only fewer ECG leads to produce meaningful diagnoses with high performance. We introduce a simple experiment to test whether contrastive learning can be applied to this task. More specifically, we added the similarity between the embedding vectors when the 12 leads signal and the fewer leads ECG signal to the loss function to bring these representations closer together. Despite its simplicity, this has been shown to have improved the performance of diagnosing with all lead combinations, proving the potential of contrastive learning on this task.

\section{Problem definition and methods} 

\begin{figure}[!ht]
  \centering
  \includegraphics[height=150pt]{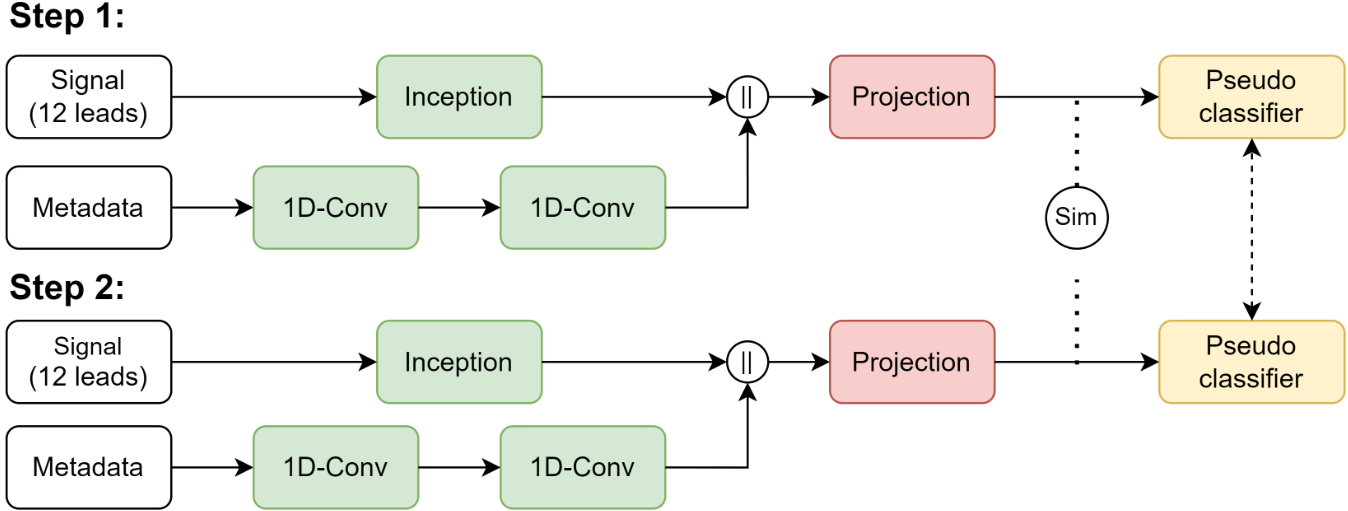}
  \caption{Illustration of the proposed approach.}
  \label{fig:main_diagram}
\end{figure}

The missing leads problem is defined in Physionet CinC2020 \cite{alday2020classification}, which requires competitors to have models that can achieve high performance on predefined sets of leads. Models tackling this task \cite{xu2022abnormality,9662737,9662723,wickramasinghe2021multi} utilized methods that focused on training and testing on the same number of leads. Here we approach this problem differently by trying to achieve higher performance on less number of leads assuming we have access to all 12 leads ECG with the use of contrastive learning.
Figure~\ref{fig:main_diagram} illustrates an overview of the experiment we used to determine whether contrastive learning would work on this task. The training process is split into two steps. In the first step, we used InceptionTime \cite{DBLP:journals/corr/abs-1909-04939} - a powerful 1D-CNN for feature extraction of 12-lead ECGs in this task. This feature is then concatenated with the encoded metadata feature before being projected onto the embedding space using a linear layer. The embedding vectors then go through a MLP (pseudo-classifier) for classification. In the second step, we froze the weights of the pseudo classifier from the first step (which maps the vectors from the embedding space to the predictions) and then trained the model with fewer leads ECG and a different CNN to project it onto the same embedding space. Notably, in this step, we bring embedding vectors of missing leads closer to their 12-lead version using the modified loss function:
\begin{equation}
\label{eq:earl}
\mathcal{L} = \mathcal{L}_{cls}(y,y^{*}) + sim(v^{12},v^{x}) \times \alpha
\end{equation}

Where \(y\) and \(y^{*}\) are the prediction and ground truth of the labels, \(v^{12}\) and \(v^{x}\) denote the projected embedding vectors of the 12-lead and its x-lead version (x  \(\epsilon\) \{2,3,4,6\}). The \(sim\) function here calculates the similarity between those two vectors and can be the L1, L2 norm of the difference between these two vectors or the cosine similarity, and \(\alpha\) is a hyperparameter. The prediction of this method is compared to that of training the model without the first step, freezing the pseudo-classifier and the custom loss function. Patients' metadata (weight, height, age, gender) is encoded using soft label encoding \cite{Cai2022,Li2020}.

\section{Experiments and Results} 

Our experiment was conducted on the PTB-XL dataset \cite{Wagner2020:ptbxlphysionet}, which contains over 21000 12-lead ECG signals, and these samples are 10 seconds long. While the dataset also provides various tasks, here we focus on the super-diagnostic classification task. We used validation Macro-AUC as the evaluation metric as the PTB-XL benchmark paper \cite{strodthoff2020deep} suggests while following the same train-validation split as the paper. Table \ref{tab:my_label} illustrates the experimental results of various leads combinations with and without the two-step training process (denoted as pseudo=True and pseudo=False). As can be seen, our method outperforms the traditional training process for all leads combination.

\begin{table}[H]
    \centering
    \small{
    \begin{tabular}{l l l}
    \hline
         \textbf{No. of leads} & \textbf{pseudo = True} & \textbf{pseudo = False} \\
         \hline
                12 & 0.9320 & N/A\\
                6  & \textcolor{red}{0.9017} & 0.9007\\
                4  & \textcolor{red}{0.9227} & 0.9218\\
                3  & \textcolor{red}{0.9213} & 0.9199\\
                2  & \textcolor{red}{0.9017} & 0.9002\\
         \hline
    \end{tabular}
    }
    \caption{Experiment results on the PTB-XL dataset}
    \label{tab:my_label}
\end{table}

\section{Discussions and Conclusion}

 We showed that contrastive learning can be used in the task of missing leads ECG classification. Despite its simplicity, the results suggest that by bringing the missing lead representation closer to that of the complete signal, diagnostic performance can be improved on all lead combinations. In the future, we will apply more complex contrastive learning techniques to improve the model’s performance.

 \newpage

\bibliographystyle{abbrv}
\bibliography{references}

\end{document}